\documentclass[letterpaper, 10 pt, conference]{ieeeconf}  % Comment this line out if you need a4paper

\IEEEoverridecommandlockouts                              % This command is only needed if 
\usepackage{xcolor} % for highlighting text
\usepackage{soul} % for highlighting text
\usepackage{amsfonts} % for double-stroke
\usepackage{amsmath} % for align
\usepackage{graphicx}
\usepackage{hyperref} % for https

\title{\LARGE \bf
	From Noise to Insight: Visualizing Neural Dynamics with Segmented SNR Topographies for Improved EEG-BCI Performance*
}

\author{Eva Guttmann-Flury$^{1,2}$, Shan Zhao$^{3}$, Jian Zhao$^{1}$, and Mohamad Sawan$^{2}$% <-this % stops a space
\thanks{*This work is supported in part by STI 2030
	Major Projects, the National Key Research and Development Program of China and the Natural Science Foundation of China (Grant No. 2022ZD0208500).}% <-this % stops a space
\thanks{$^{1,2}$Eva Guttmann-Flury is with the Department of Micro-Nano Electronics and the MoE Key Laboratory of Artificial Intelligence, 
	Shanghai Jiao Tong University, Shanghai, China,
	and visiting at CenBRAIN Neurotech, School of Engineering,  
	Westlake University, Hangzhou, China
        {\tt\small eva.guttmann.flury@gmail.com}}%
\thanks{$^{3}$Shan Zhao is with the School of Public Health, Shanghai Jiao Tong University School of Medicine, Shanghai, China,
        {\tt\small shanzhao23@sjtu.edu.cn}}%
\thanks{$^{1}$Jian Zhao is with the Department of Micro-Nano Electronics and the MoE Key Laboratory of Artificial Intelligence, 
	Shanghai Jiao Tong University, Shanghai, China
{\tt\small zhaojianycc@sjtu.edu.cn}}%
\thanks{$^{2}$Mohamad Sawan is with the CenBRAIN Neurotech, School of Engineering,  
Westlake University, Hangzhou, China
{\tt\small sawan@westlake.edu.cn}}%
}

\begin{document}

\maketitle
\thispagestyle{empty}
\pagestyle{empty}

%%%%%%%%%%%%%%%%%%%%%%%%%%%%%%%%%%%%%%%%%%%%%%%%%%%%%%%%%%%%%%%%%%%%%%%%%%%%%%%%
\begin{abstract}

Electroencephalography (EEG)-based wearable brain-computer interfaces (BCIs) face challenges due to low signal-to-noise ratio (SNR) and non-stationary neural activity. We introduce in this manuscript a mathematically rigorous framework that combines data-driven noise interval evaluation with advanced SNR visualization to address these limitations. Analysis of the publicly available Eye-BCI multimodal dataset demonstrates the method's ability to recover canonical P300 characteristics across frequency bands (delta: 0.5-4 Hz, theta: 4-7.5 Hz, broadband: 1-15 Hz), with precise spatiotemporal localization of both P3a (frontocentral) and P3b (parietal) subcomponents. To the best of our knowledge, this is the first study to systematically assess the impact of noise interval selection on EEG signal quality. Cross-session correlations for four different choices of noise intervals spanning from early to late pre-stimulus phases also indicate that alertness and task engagement states modulate noise interval sensitivity, suggesting broader applications for adaptive BCI systems. While validated in healthy participants, our results represent a first step towards providing clinicians with an interpretable tool for detecting neurophysiological abnormalities and provides quantifiable metrics for system optimization.

\end{abstract}

%%%%%%%%%%%%%%%%%%%%%%%%%%%%%%%%%%%%%%%%%%%%%%%%%%%%%%%%%%%%%%%%%%%%%%%%%%%%%%%%
\section{INTRODUCTION}
% 1. Brain-Computer Interfaces 
Brain-computer interfaces (BCIs) enable direct interaction between neural activity and external devices by decoding cortical signals without relying on the body's peripheral motor system. This framework enables users to control assistive technologies, such as prosthetics or communication aids, and facilitates neural monitoring and clinical diagnostics. Through real-time or offline decoding of brain signals, BCIs can detect user intent as well as neurophysiological biomarkers relevant to neurological conditions. This dual functionality highlights their broad utility in both medical and research contexts. BCIs thus hold potential to transform both treatment and understanding of medical disorders \cite{Mridha_2021}.

% 2. EEG
Among various neural recording modalities, electroencephalography (EEG) has become the predominant choice for non-invasive BCIs due to its excellent temporal resolution, portability, and cost-effectiveness. Despite these advantages, EEG signals are often contaminated by noise and exhibit non-stationary behavior, presenting significant challenges for reliable decoding and requiring advanced processing techniques \cite{Peksa_2023}.

% 3. Common Signals for EEG-Based BCIs: P300 
The P300 event-related potential (ERP) has become a cornerstone of EEG-based BCI research due to its robust and stimulus-locked nature that requires no explicit user training. The P300 is typically divided into two subcomponents: the P3a and P3b. The P3a emerges over frontocentral regions around 250 -- 280 ms and is associated with automatic attention responses to novel stimuli. The P3b, typically peaking between 300 -- 500 ms over parietal regions, reflects cognitive evaluation of task-relevant stimuli and is prominently used in paradigms such as BCI spellers. P300 characteristics are highly sensitive to cognitive load, task complexity, and subject condition. Importantly, reduced P3b amplitude and prolonged latency serve as potential biomarkers for cognitive deterioration, enabling the distinction between healthy individuals, those with mild cognitive impairment (MCI) and those with Alzheimer’s disease (AD) \cite{Hedges_2016, Morrison_2019}.

% 4. Challenges of EEG-Based BCIs: Noise and Variability
Despite its distinct neural signature, the P300 is challenging to isolate reliably due to noise and variability in EEG signals. Neural responses recorded via EEG are often low in amplitude and obscured by ongoing background brain activity, rendering them vulnerable to both biological and technical artifacts. Inter-subject differences, as well as intra-subject variations across sessions, further hinder consistent signal interpretation, necessitating extensive per-user calibration sessions and frequent recalibration. Endogenous factors — such as attention lapses, fatigue, and dynamic cognitive states — introduce non-stationarity, while exogenous sources like eye movements, muscle activity, and electrode interference can significantly corrupt signal quality. Even state-of-the-art machine learning methods frequently fail to generalize across such variability, compromising real-world reliability and scalability of EEG-based BCIs \cite{Chaddad_2023, Islam_2016}.

% 5. Signal-to-Noise Ratio
The signal-to-noise ratio (SNR) serves as a critical metric for evaluating and optimizing BCI performance. In P300 systems, higher SNR values correlate strongly with improved oddball detection accuracy and reduced trial requirements \cite{Artzi_2018}. However, conventional methods for calculating SNR often rely on arbitrarily defined noise intervals. These intervals may inadvertently include task-related neural activity or fail to account for the non-stationary nature of background noise. This limitation becomes particularly acute in P300 paradigms, where standard pre-stimulus baselines, typically starting between -1000 ms to -100 ms, may overlap with anticipatory potentials or habituation effects \cite{Kota_2005, Struber_2002, Samima_2017}.

\begin{figure}[!htp]
	\centering
	\includegraphics[width=7cm]{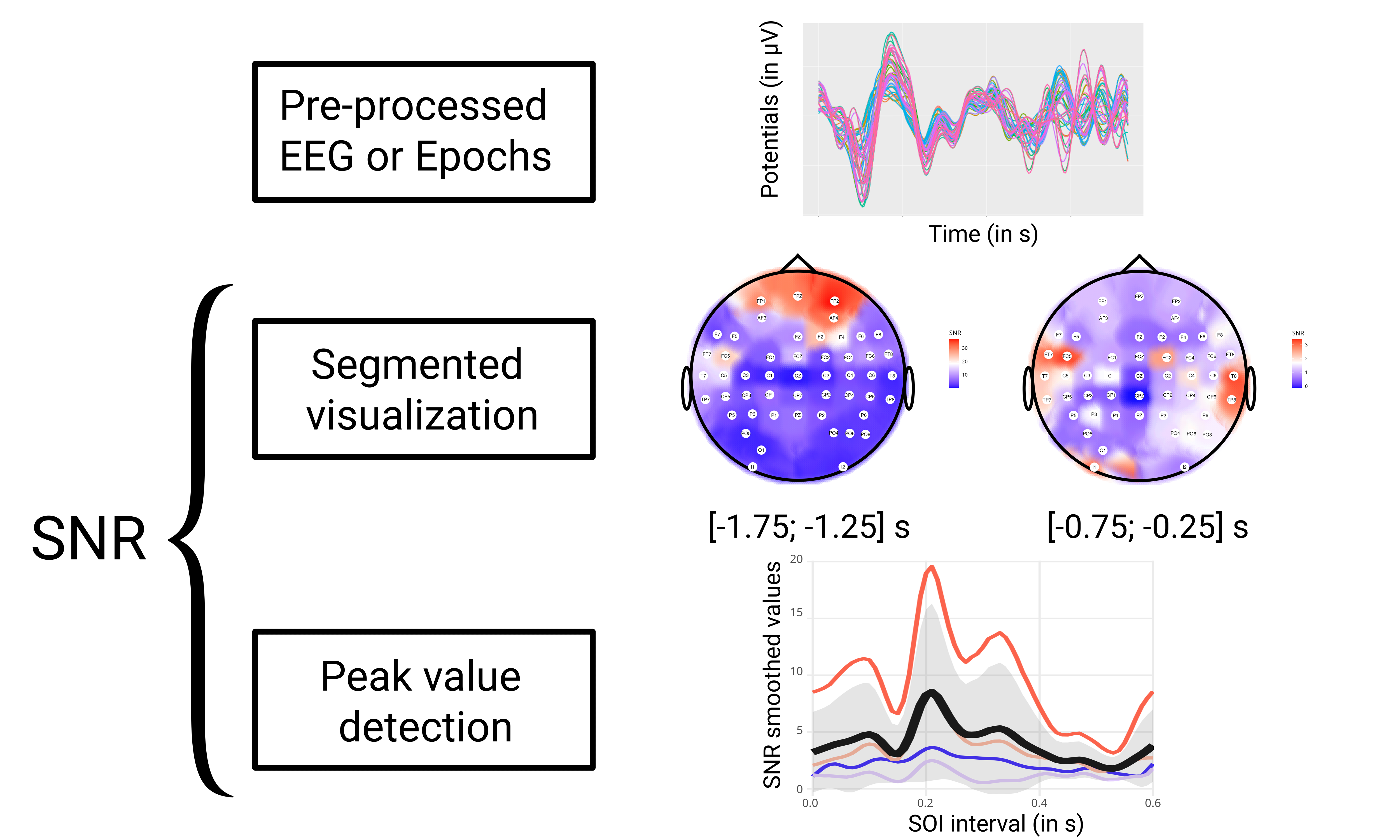}
	\caption
	[Goal]
	{Flowchart including the SNR framework, highlighting the impact of noise range selection} 
	\label{fig:Goal}
\end{figure}

% 6. The Role of Visualization in Enhancing Interpretability and Performance
To address these challenges and improve interpretability, we propose a novel framework that combines data-driven noise interval selection with advanced visualization techniques. Our approach systematically evaluates different pre-stimulus intervals spanning from early to late pre-stimulus phases ([-1.75, -1.25]s, [-1.1, -0.6]s, [-0.75, -0.25]s, and [-0.3, 0]s) to empirically determine their impact on P300 SNR characteristics, as illustrated by the divergent spatial patterns in simultaneous segmented SNR topographies in Figure \ref{fig:Goal}. Furthermore, our method provides spatiotemporal SNR mapping across scalp regions, bridging the gap between signal processing and machine learning approaches through interpretable visualization.

% 7. Contributions
This work makes two key contributions to the field of EEG-based BCIs: (1) a novel framework for assessing SNR spatial distribution to enhance interpretability of spatiotemporal dynamics and (2) a systematic evaluation of noise interval selection on SNR characteristics to replace arbitrary noise intervals. By establishing data-driven approaches to noise definition and visualization, our framework advances both the technical capabilities and interpretability of EEG-based BCIs, addressing critical limitations in current implementations. These contributions support the development of more interpretable and reliable BCI systems, particularly in noisy and variable recording environments.

\section{METHODS}
The methodological SNR framework of this study builds upon the non-stationary and nonlinear nature of large-scale neural dynamics. During cognitive task performance, synchronized neuronal activity exhibits time-varying statistical properties — both in mean activation patterns and cross-channel covariance structures. These characteristics are believed to reflect the brain's self-organizing dynamics \cite{Schoner_1988}, where evolving functional connectivity between cortical regions generates the non-Gaussian signal distributions observed in EEG recordings. Such intrinsic variability directly impacts the signal quality challenges, motivating our data-driven approach to noise characterization and SNR optimization.

\subsection{Data Description}
We analyzed EEG recordings from the publicly available Eye-BCI multimodal dataset, hosted on Synapse \cite{eye_bci_multi_dataset}, which comprises 63 experimental sessions from 31 neurologically intact participants (age range: 22–57 years; 2 left-handed) and includes 2,520 P300 trials \cite{GuttmannFlury_Dataset_2025}. Participants provided standardized self-reports of alertness levels and task engagement at scheduled breaks, enabling quantitative assessment of vigilance states during recordings. Fifteen subjects completed three repeated sessions, yielding longitudinal data for stability analysis. The sample size was determined through rigorous statistical power analysis using the Fitted Distribution Monte Carlo (FDMC) simulation method \cite{GuttmannFlury_APriori_2019}. Code for reproducing the experimental setup is accessible at \href{https://github.com/QinXinlan/EEG-experiment-to-understand-differences-in-blinking/}{https://github.com/QinXinlan/EEG-experiment-to-understand-differences-in-blinking/}.

\subsection{Noise Characteristics}

Let's consider the measured EEG time series $x$ recorded for an EEG electrode at each time point $t$, symbolically represented as $x = x(t)$. The measured signal $x$ consists of two fundamental components: the signal of interest (SOI) $s$ representing task-relevant neural activity, and additive noise $n^a$ encompassing all other signal components: $x = s + n^a$. The SOI emerges when specific cortical sources become activated in response to experimental stimuli or cognitive tasks, while the additive noise includes both physiological artifacts and measurement noise.

The noise component can be further decomposed into four distinct categories. First, the basic noise $n^b$ represents ever-present background activity unrelated to the experimental paradigm. Second, event-generated added noise includes neural activity specifically induced by the task but unrelated to the SOI. Third, subtracted noise comprises neural activity suppressed during task performance. Fourth, signal variance-generated noise accounts for trial-to-trial variability in SOI characteristics. This comprehensive noise model acknowledges that activated cortical sources may appear at varying latencies and intensities across trials, with multiple neural processes becoming engaged or suppressed during task performance.

The temporal dynamics of cortical activation exhibit inherent variability across trials, with both the latency and amplitude of neural responses fluctuating significantly. During task performance, multiple cortical regions typically become engaged in parallel with the SOI, creating a complex activation pattern where some neural processes are specifically enhanced by the task (constituting added noise) while others are simultaneously suppressed (representing subtracted noise). These complementary event-related noise components are collectively designated as $n^e$ in our framework.

While the SOI can be identified despite these temporal variations, its magnitude remains subject to trial-to-trial fluctuations. The conventional modeling approach assumes statistical independence between the SOI and background neural activity, implying that the SOI component should theoretically reduce to zero in the absence of task-related events. This fundamental assumption underlies our baseline correction procedures and noise estimation methodology.

To enable meaningful comparison across epochs, signals are typically centered by subtracting the mean activity during a predefined baseline period (e.g., between $t_1 = -100 \hspace{0.3em} ms$ to $t_2 = 0 \hspace{0.3em} ms \hspace{0.1em}$) \cite{Graben_2000}. The baseline-corrected measured signal at an epoch is given by:
\begin{equation}
	x_i^c = s_i + n_i^b + n_i^e = \overline s + n_i
\end{equation}
where $\overline s$ is the mean of the SOI. For simplicity, we refer to the baseline-corrected signal as $x$. The complete noise during an epoch combines all components:
\begin{equation}
	n_i = s_i - \overline s + n_i^b + n_i^e = n_i^b + n_i^s 
\end{equation}
The centered basic noise $n^b$ is assumed to be stationary and ergodic, represented by its variance since its mean is zero. The event-generated noises $n^e$, being time-dependent, cannot be assumed ergodic. The signal variance-generated noise similarly fails to meet ergodicity assumptions due to its inherent variability. Several factors contribute to this non-stationary characteristic: task habituation effects may systematically alter SOI intensity across the experiment, minor electrode displacement can introduce measurement inconsistencies, and numerous other uncontrolled variables may further modulate the signal properties. 

Signal analysis approaches must account for these complex noise characteristics. The peak measurement method proves effective for prominent neural responses identifiable in single trials and ERPs. For smaller effects, the area under the curve is preferred \cite{Graben_2001}. We distinguish between theoretical (double-struck: expected mean $\mathbb{E}$ and variance $\mathbb{V}$) and practical (single-struck: $E$ and $V$) estimates of moments. The sample mean is an unbiased estimator of the population mean, which can be computed with:
\begin{equation}
	E(x) = \frac{1}{N} \sum_{i=1}^N \frac{1}{T}  \int_0^T x_i(t) 
\end{equation}
And the unbiased variance is:
\begin{equation}
	V(x) = \frac{1}{N-1} \sum_{i=1}^N \frac{1}{T} \int_0^T (x_i(t) - {\overline x} (t) )^2
	\label{eq:Var}
\end{equation}

\subsection{Signal-to-Noise Ratio}
The signal-to-noise ratio (SNR) is defined as the dimensionless ratio of signal power to noise power. In EEG analysis, we distinguish between single-trial $SNR_T$ and grand average $SNR_{GA}$. The grand average of event-related potentials (ERPs) typically demonstrates superior SNR characteristics, as signal averaging enhances the consistent neural response while suppressing random noise components through phase cancellation.

The practical significance of SNR becomes evident when examining neural patterns embedded within background EEG activity. Low SNR conditions, where target signals are obscured by physiological noise, present substantial detection challenges and correspond to small effect sizes. Conversely, high SNR scenarios facilitate robust signal detection and classification in BCI systems. This relationship underscores the critical importance of SNR optimization for reliable single-trial analysis in both research and clinical applications.

The mathematical formulation begins with the single-trial signal decomposition: $x_i = \overline s + n_i$ as the sum of the mean SOI and the whole noise, or equivalently the sum of the (varying) SOI and the centered basic noise $x_i = s_i + n_i^b$. 

Two key assumptions enable subsequent derivations: (1) the mean SOI remains constant across trials, and (2) the centered baseline noise $n_i^b$ is both stationary and ergodic. These properties ensure statistical independence between the signal and noise components. The power of the measured signal, defined as the expectation of its squared magnitude, follows from these assumptions:
\begin{align}
	\mathbb{P}(x_i) & = \mathbb{E}[x_i^2] = \mathbb{E}[s_i^2 + 2 * s_i * n_i^b + (n_i^b)^2] \nonumber \\
	& = \mathbb{E}[s_i^2]  + 2 * \mathbb{E}[s_i] * \mathbb{E}[n_i^b] + \mathbb{E}[(n_i^b)^2]  
\end{align}
Since the mean of the basic noise is assumed to be equal to zero, $\mathbb{E}[n_i^b] = \overline {n^b} = 0$, the power of the measured signal is simply:
\begin{equation}
	\mathbb{P}(x_i)  = \mathbb{E}[s_i^2]  + \mathbb{E}[(n_i^b)^2] = \mathbb{P}(s_i) + \mathbb{P}(n_i^b) 
\end{equation}
between the signal power $\mathbb{P}(s_i)$ and the basic noise power $\mathbb{P}(n_i^b)$:
\begin{align}
	\mathbb{SNR_T} = \frac{\mathbb{P}(n_T^{signal})}{\mathbb{P}(n_T^{noise})} = \frac{\mathbb{P}(s_i)}{\mathbb{P}(n_i^b)} =  \frac{\mathbb{P}(x_i) - \mathbb{P}(n_i^b) }{\mathbb{P}(n_i^b)} \nonumber \\
	\mathbb = \frac{\mathbb{P}(x_i) }{\mathbb{P}(n_i^b)} - 1
\end{align}

The stationary and ergodic properties of the centered basic noise $n^b$ enable flexible estimation of its power characteristics. Since these assumptions imply time-invariant statistical properties, the noise power can be reliably computed from any sufficiently long signal segment where the SOI is absent. For practical calculation, we define:
\begin{equation}
	SNR_T = \frac{ \frac{1}{N} \sum_{i=1}^N \frac{1}{\Delta t_s} \int_{t_s}^{t_s+\Delta t_s} x_i^2(t) }{ \frac{1}{N} \sum_{i=1}^N \frac{1}{\Delta t_n} \int_{t_n}^{t_n+\Delta t_n} x_i^2(t) } - 1 
	\label{eq:SNRt}
\end{equation}
where $t_n$ (resp. $t_s$) marks the onset of the noise (resp. signal) estimation window, $\Delta t_n$ (resp. $\Delta t_s$) represents the integration duration, and $N$ is the number of trials.

\section{RESULTS}
To validate the efficiency of our segmented SNR topography framework, we first demonstrate its ability to recover well-established neural signatures before extending the analysis to noise interval optimization. In this paper, we focus on the P300 response — a robust and temporally precise ERP component — to verify that our method accurately captures its characteristic latency and spatial distribution. 

With these foundational validations in place, we then leverage SNR dynamics to systematically evaluate noise interval definitions, establishing data-driven criteria for baseline selection that account for temporal non-stationarities. This hierarchical presentation ensures that our framework’s sensitivity to neural patterns is rigorously benchmarked against known physiological principles before addressing its novel applications.

\subsection{Signal Preprocessing and SNR Computation} 
All EEG data underwent a standardized preprocessing pipeline to ensure robust signal quality prior to SNR analysis. The raw EEG data were first processed using the Adaptive Blink and Correction De-drifting (ABCD) algorithm for bad channel detection and blink removal \cite{GuttmannFlury_ABC_2019}. Signals were then re-referenced using a surface Laplacian filter to enhance spatial resolution by attenuating volume-conducted artifacts while emphasizing local cortical activity \cite{Kayser_2015}. 

The preprocessed data were then analyzed across multiple spectral configurations to assess their impact on signal quality. The unfiltered signal provided the baseline measurement for comparison. Three frequency bands of established relevance to P300 responses were then examined using a 4th-order Butterworth filter: the conventional 1-15 Hz range encompassing typical P300 components, along with two specific oscillatory bands - delta (0.5-4 Hz) and theta (4-7.5 Hz) - known to reflect distinct neural processes underlying oddball target detection. This selection was motivated by extensive evidence linking delta activity to P300 amplitude modulations and theta oscillations to working memory engagement during target discrimination tasks, particularly in prefrontal-parietal networks \cite{Başar-Eroglu_2001, Bougrain_2012}.

\subsection{Spatiotemporal Visualization of the P300 ERP} 
\begin{figure}[!htp]
	\centering
	\includegraphics[width=7cm]{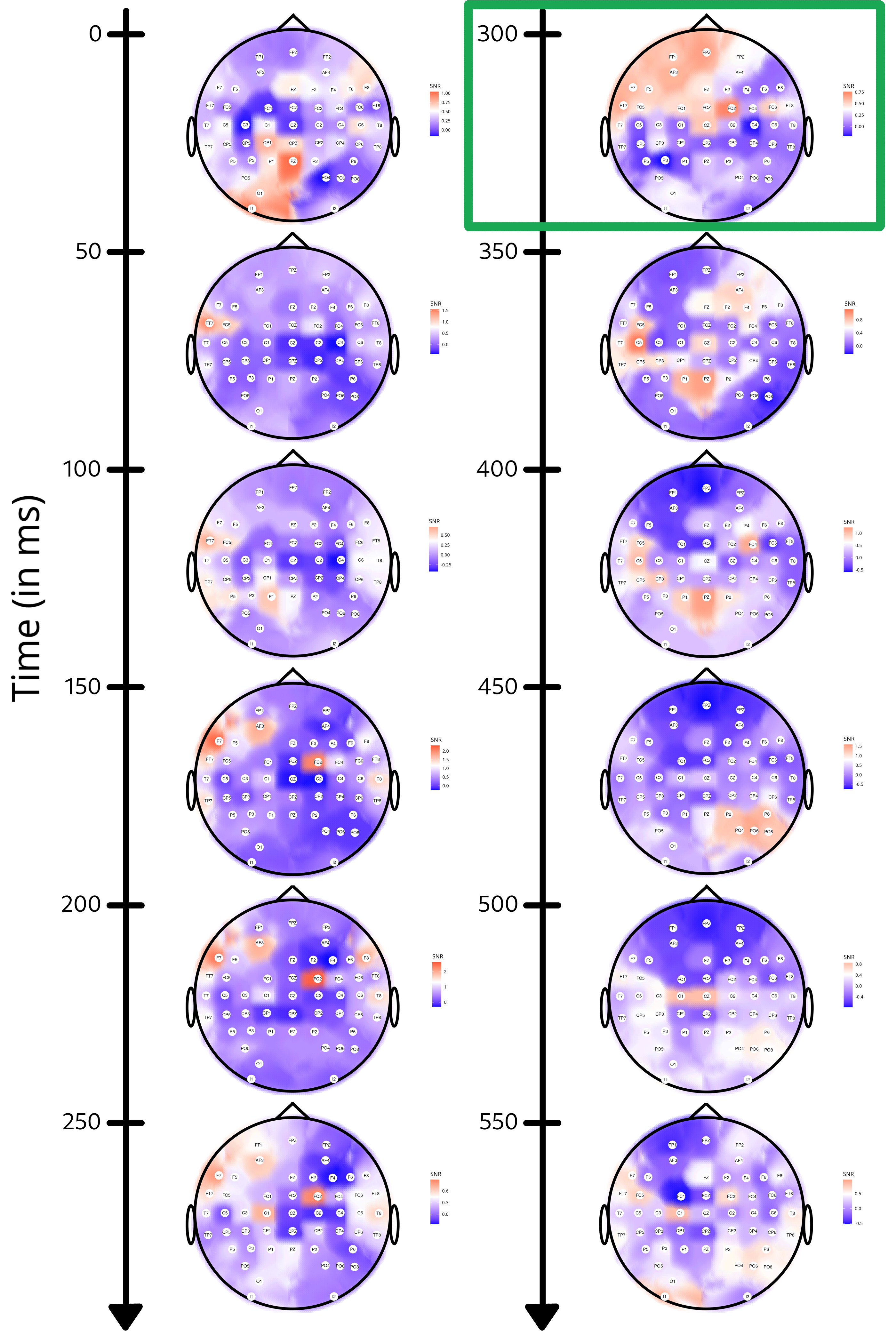}
	\caption
	[P300]
	{Segmented SNR topographies of the P300 for a specific subject in the 1-15 Hz frequency range}
	\label{fig:P300}
\end{figure}

Given the well-characterized temporal dynamics of the P300 response, we first examined the segmented SNR topography across predefined time intervals. Figure \ref{fig:P300} demonstrates clear spatiotemporal evolution of signal quality, with SNR peaks corresponding precisely to the expected 300 ms post-stimulus window of the P300 component for the displayed subject in the 1-15 Hz frequency range. 
 
Beyond the segmented SNR topographies, the time-resolved SNR profiles (computed with a 0.01 s intervals for the SOI) underwent smoothing via a 20 ms moving average to attenuate high-frequency fluctuations while preserving P300-relevant dynamics.  An adaptive thresholding procedure was also implemented to enhance spatial specificity, retaining only electrodes exceeding $mean(SNR(t))+k*sd(SNR(t))$ at each time point $t$. 
In this formulation, $mean(SNR(t))$ and $sd(SNR(t))$ represent the cross-electrode average and standard deviation of SNR values at time $t$. The selectivity parameter $k$ (empirically set at 2) governs the trade-off between spatial focus (elevated $k$) and sensitivity to lower-amplitude components (reduced $k$).

\begin{figure}[!htp]
 	\centering
 	\includegraphics[width=5cm]{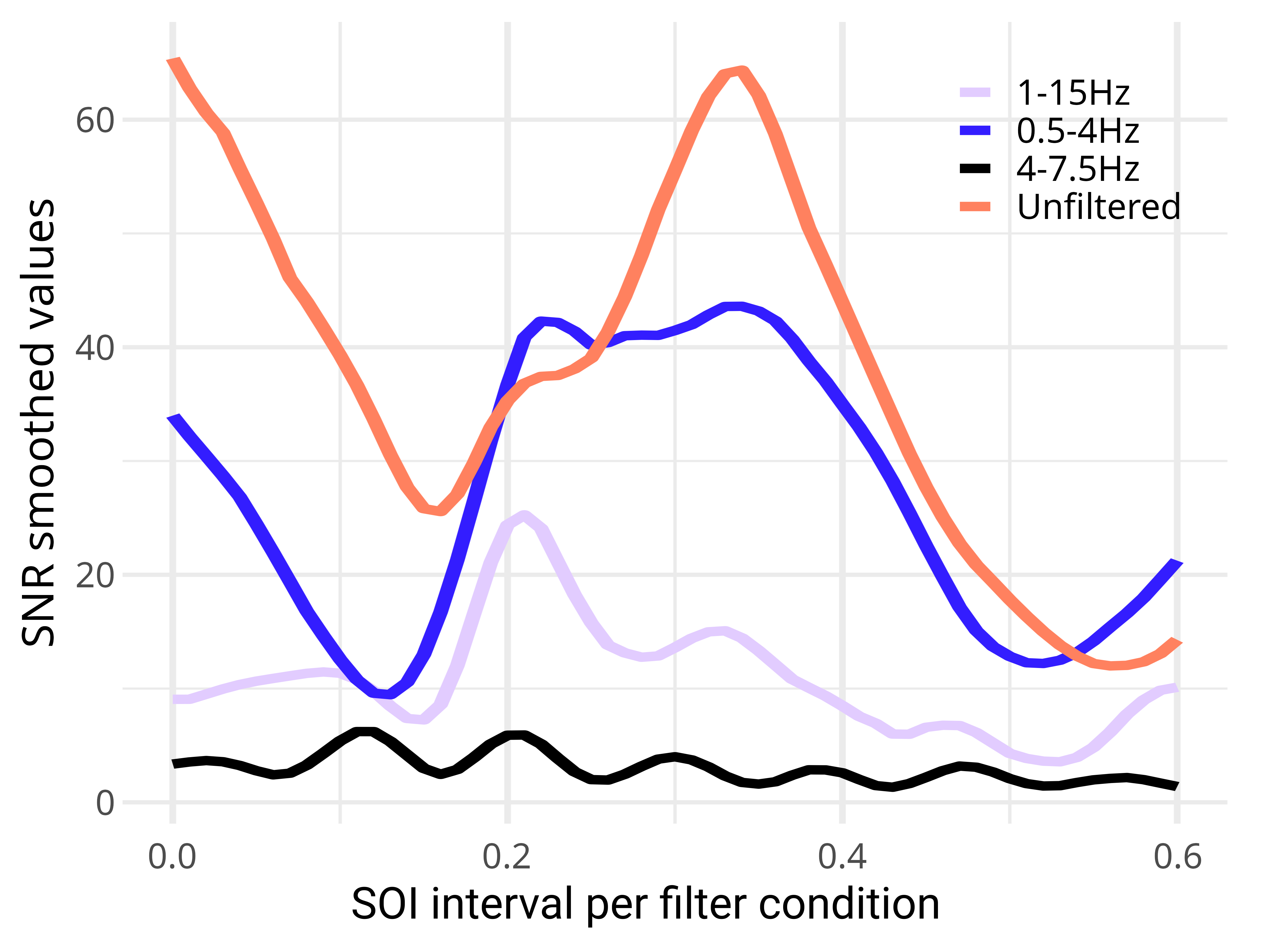}
 	\caption
 	[DiffFreq]
 	{Single-subject SNR smoothed values per filter type revealing preserved temporal precision for P3a and P3b}
 	\label{fig:DiffFreq}
 \end{figure}

Figure \ref{fig:DiffFreq} illustrates the frequency-specific SNR patterns for a subject, demonstrating the framework's ability to resolve distinct P300 subcomponents. Single-subject analysis preserves temporal precision that would be obscured in grand averages due to inter-subject latency variability.
%demonstrates this approach's efficiency across frequency conditions for a subject, revealing differential enhancement of P300 subcomponents. 
The P3a emerges most prominently in both broadband (1-15 Hz) and theta band (4-7.5 Hz) analyses at frontocentral electrodes, consistent with its association with novelty detection and early attentional engagement. The P3b subcomponent dominates unfiltered signals  - with secondary expression in delta-band (0.5-4 Hz) - over parietal regions, which suggests its emergence from integrated multi-frequency activity during later cognitive evaluation. 
 
\subsection{Noise Interval Optimization Effects}

\begin{figure}[!htp]
	\centering
	\includegraphics[width=8.5cm]{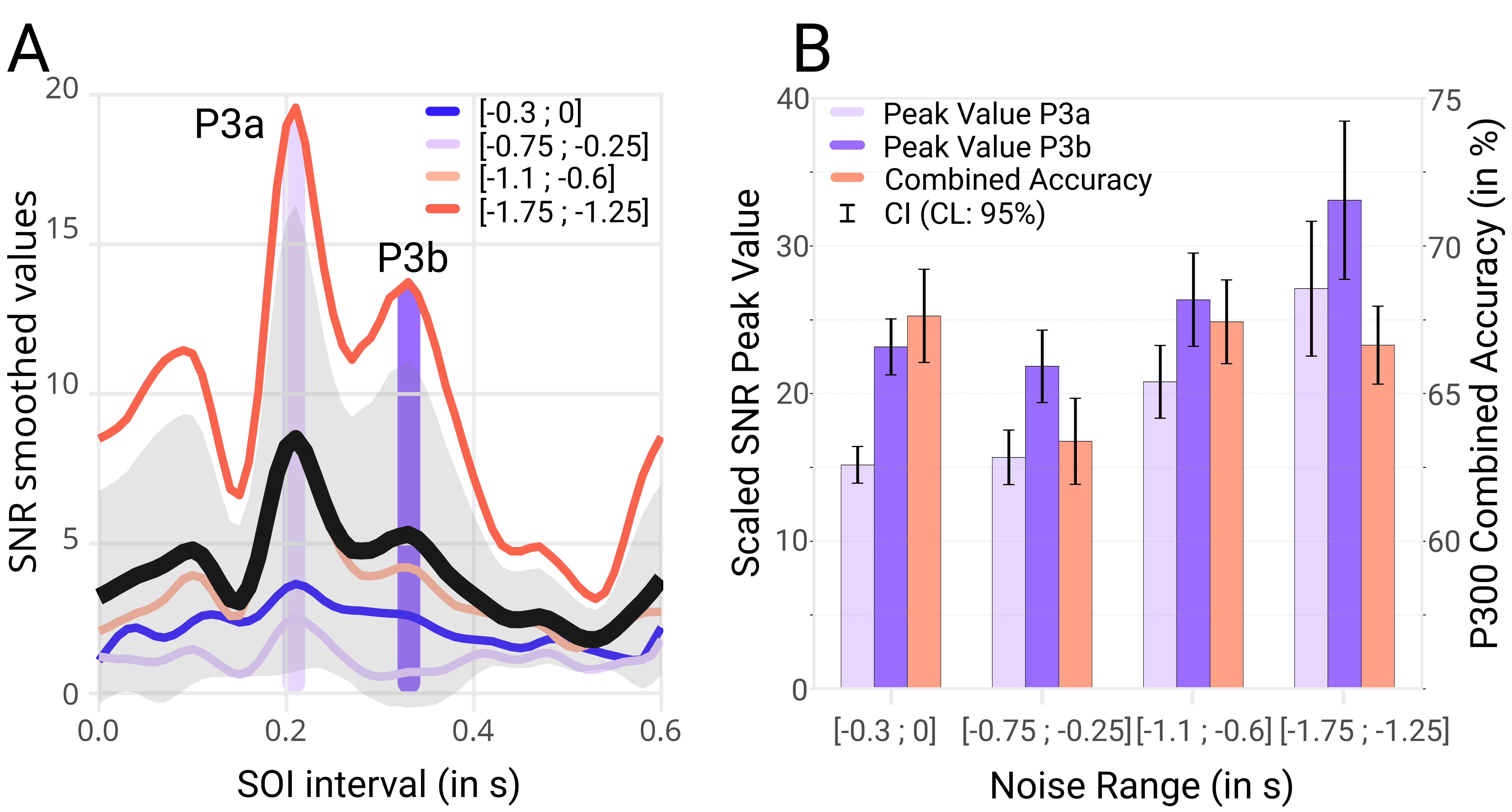}
	\caption
	[DiffNoise]
	{A. SNR smoothed values per noise range for a subject (mean in black and standard deviation in grey); B. P3a and P3b mean scaled SNR peak value for all 63 sessions (31 subjects)}
	\label{fig:DiffNoise}
\end{figure}

The systematic evaluation of noise intervals reveals distinct effects on SNR profiles and P300 subcomponent visibility. Analysis across all 63 sessions utilizes a standardized source reconstruction pipeline where each epoch undergoes independent processing using an average head model. The pipeline incorporates noise-prewhitening of leadfield matrices through regularized covariance estimation, effectively minimizing spurious correlations between amplitude estimates and sample size. Spatial averaging across 75 FreeSurfer-defined cortical parcels yields reconstructed activity that maintains anatomical specificity while optimizing computational efficiency.
Temporal isolation of P300 subcomponents employs dedicated analysis windows: 150--250 ms for P3a and 300--500 ms for P3b.
The resulting activation patterns (Figure \ref{fig:DiffNoise}B) – quantified through the combination of P3a and/or P3b peak values for classification – systematically replicate interval-dependent SNR relationships observed in single-subject SNR values (Figure \ref{fig:DiffNoise}A).
% The resulting regional activation patterns, represented by P3a and P3b peak values (Figure} \ref{fig:DiffNoise}\hl{B), which combination allow for the classification, consistently reflect the interval-dependent relationships observed in single-subject SNR values (Figure} \ref{fig:DiffNoise}\hl{A). 
Earlier baselines preferentially enhance P3b detection, while later intervals provide more systematic capture of the full subcomponent spectrum, as indicated by reduced confidence intervals.

%The systematic evaluation of noise intervals revealed significant impacts on SNR profiles and consequent P300 subcomponents visibility, as illustrated in Figure \ref{fig:DiffNoise}. Earlier baseline intervals ([-1.75; -1.25]s) produced superior SNR enhancement for the P3b component in parietal regions, likely by excluding anticipatory potentials that contaminate conventional pre-stimulus periods. Conversely, later intervals ([-0.3; 0]s and [-0.75; -0.25]s) artificially attenuated P3b detection while paradoxically improving P3a resolution at frontocentral sites, suggesting differential sensitivity to baseline correction artifacts across subcomponents. These findings establish that noise interval selection non-uniformly filters neural dynamics, with earlier intervals preferentially preserving late cognitive components (P3b) and intermediate windows capturing full subcomponent spectra.

\begin{figure}[!htp]
	\centering
	\includegraphics[width=5cm]{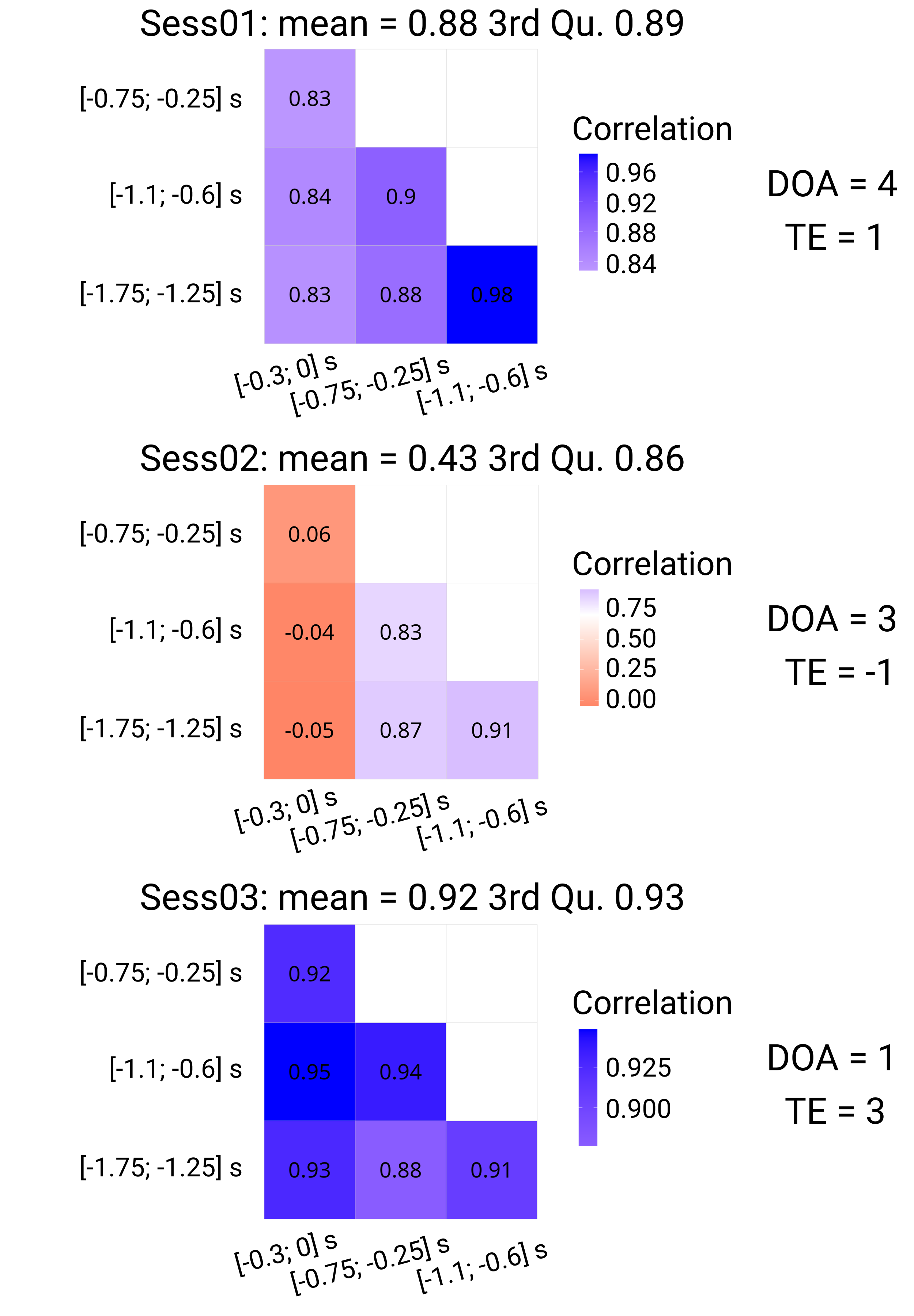}
	\caption
	[Corr]
	{Noise range correlations (Kendall’s $\tau$) for a subject, associated with Degree Of Alertness (DOA: lower = rested, higher = fatigued) and Task Engagement (TE: lower = disengaged, higher = engaged) }
	\label{fig:Corr}
\end{figure}

To quantify the temporal stability of noise interval effects, cross-session correlations (Kendall’s $\tau$) are computed between SNR profiles generated from different noise intervals. Figure \ref{fig:Corr} presents the resulting heatmap correlation matrices for a subject across three sessions. An intriguing association is observed between self-reported alertness levels and the stability of interval-wise SNR correlations, suggesting that attentional state modulates noise interval sensitivity. While these subjective reports inherently lack the precision of physiological measures, the identified trend aligns with neurocognitive expectations: during periods of either extreme fatigue (low alertness) or hyper-focus (high attention), correlation matrices show greater uniformity across noise intervals, indicating reduced dependence on baseline selection when neural responses approach maximum or minimum amplitude thresholds. This pattern implies that noise interval optimization becomes most critical for intermediate vigilance states, where neural signals exhibit sufficient variability to benefit from data-driven baseline correction.

\section{DISCUSSION}

\subsection{ Visualization Framework for Enhanced Interpretability}
The proposed SNR visualization framework provides a principled approach to analyzing neural signal quality across spatial, temporal, and spectral domains. By representing SNR variations as topographic maps, the method reveals key features of P300 subcomponents — including the frontocentral P3a and parietal P3b. The thresholded SNR profiles offer quantitative insights into signal stability while maintaining physiological interpretability, addressing a critical need in both basic research and clinical applications.

The framework's strength lies in its ability to make non-stationary EEG dynamics intuitively comprehensible while maintaining mathematical rigor. Such visualizations are particularly valuable for translational applications, allowing clinicians to rapidly identify pathological patterns (e.g., attenuated parietal SNR in MCI patients) and engineers to optimize system parameters based on spatially localized signal quality metrics. By representing SNR variations through intuitive color gradients and temporal profiles, the method facilitates rapid assessment of signal quality without requiring advanced signal processing expertise — a feature especially important for clinical deployment.

\subsection{Alertness-Dependent SNR Stability}
The observed link between vigilance states and noise interval correlations suggests broader applications beyond methodological optimization. First, the reduced sensitivity to baseline selection during extreme alertness states (both hyper-focus and fatigue) implies that adaptive BCI systems could simplify signal processing in these conditions — potentially enabling more efficient operation during critical tasks. Second, the intermediate vigilance states where interval choice matters most may represent an optimal "sweet spot" for cognitive monitoring, as they reflect natural fluctuations in attention rather than ceiling/floor effects.

An alternative interpretation emerges when considering clinical populations: the greater uniformity of correlations during fatigue states could reflect pathological attenuation of neural variability in disorders like ADHD or depression, rather than simply signal saturation. This raises the possibility of using interval sensitivity itself as a marker of intact neuromodulatory function.

\subsection{Limitations}
While this study provides novel insights into SNR optimization and visualization, several limitations should be considered. First, the reliance on predefined frequency bands and noise intervals may not fully capture individual variability in neural dynamics. The generalizability of the proposed framework to clinical populations remains to be established, particularly for disorders with altered ERP morphology (e.g., schizophrenia or traumatic brain injury). Second, while the visualization framework enhances interpretability, it inherits the inherent limitations of scalp EEG, including volume conduction effects that may obscure deeper generators. Third, the alertness correlations were derived from subjective reports rather than objective physiological measures, which may introduce bias. Future work should integrate multimodal monitoring (e.g., pupillometry or skin conductance) to better characterize the relationship between vigilance states and SNR stability. Additionally, the current implementation requires offline processing; real-time adaptation of noise intervals based on continuous SNR estimation presents an important technical challenge for translational applications. 

\section{CONCLUSIONS}

This study established a systematic framework for optimizing and visualizing P300 signal quality through data-driven noise interval selection and spectrally resolved SNR analysis. The results illustrate that different noise intervals can capture both P3a and P3b subcomponents, while alertness states modulate the sensitivity to interval selection — a finding with implications for adaptive BCI design.

The practical significance lies in the framework’s dual utility: it provides clinicians with an interpretable tool for detecting neurophysiological abnormalities (e.g., attenuated parietal SNR in MCI) while offering engineers quantifiable metrics for system optimization. The methodological approach generalizes to any ERP-based paradigm, though external validity for clinical populations requires verification through larger, multisite studies.

These preliminary findings, derived from healthy participants, underscore the need for future validation across three key dimensions: (1) expanded cohorts encompassing neurological disorders, (2) standardization of real-time implementation, and (3) integration with physiological vigilance monitoring. Such systematic replication will determine whether interval-sensitive SNR profiles can transition from a laboratory tool to a clinically actionable biomarker.

\addtolength{\textheight}{-12cm}   % This command serves to balance the column lengths

%%%%%%%%%%%%%%%%%%%%%%%%%%%%%%%%%%%%%%%%%%%%%%%%%%%%%%%%%%%%%%%%%%%%%%%%%%%%%%%%

%%%%%%%%%%%%%%%%%%%%%%%%%%%%%%%%%%%%%%%%%%%%%%%%%%%%%%%%%%%%%%%%%%%%%%%%%%%%%%%%

%%%%%%%%%%%%%%%%%%%%%%%%%%%%%%%%%%%%%%%%%%%%%%%%%%%%%%%%%%%%%%%%%%%%%%%%%%%%%%%%
%\section*{APPENDIX}


%\section*{ACKNOWLEDGMENT}

%The preferred spelling of the word ``acknowledgment'' in America is without an ``e'' after the ``g''. Avoid the stilted expression, ``One of us (R. B. G.) thanks . . .''  Instead, try ``R. B. G. thanks''. Put sponsor acknowledgments in the unnumbered footnote on the first page.

%%%%%%%%%%%%%%%%%%%%%%%%%%%%%%%%%%%%%%%%%%%%%%%%%%%%%%%%%%%%%%%%%%%%%%%%%%%%%%%%

%References are important to the reader; therefore, each citation must be complete and correct. If at all possible, references should be commonly available publications.

\end{document}